\documentclass[prl,aps,floats,superscriptaddress,amsfonts,twocolumn,showpacs]{revtex4}

\usepackage{graphicx}

\begin{document}

\newcommand{\ket} [1] {\vert #1 \rangle}
\newcommand{\bra} [1] {\langle #1 \vert}
\newcommand{\braket}[2]{\langle #1 | #2 \rangle}
\newcommand{\proj}[1]{\ket{#1}\bra{#1}}
\newcommand{\mean}[1]{\langle #1 \rangle}
\newcommand{\opnorm}[1]{|\!|\!|#1|\!|\!|_2}

\title{Reverse Coherent Information}

\author{Ra\'{u}l Garc\'{\i}a-Patr\'{o}n}
\affiliation{Research Laboratory of Electronics, MIT, Cambridge, MA 02139}

\author{Stefano Pirandola}
\affiliation{Research Laboratory of Electronics, MIT, Cambridge, MA 02139}

\author{Seth Lloyd}
\affiliation{Research Laboratory of Electronics, MIT, Cambridge, MA 02139}

\author{Jeffrey H. Shapiro}
\affiliation{Research Laboratory of Electronics, MIT, Cambridge, MA 02139}

\begin{abstract}
In this letter we define a family of entanglement distribution protocols assisted by feedback classical communication
that gives an operational interpretation to reverse coherent information, i.e., the
symmetric counterpart of the well known coherent information.
This lead to the definition of a new entanglement distribution capacity that exceeds
the unassisted capacity for some interesting channels.
\end{abstract}


\maketitle

Shannon's great result was proving that sending information
through a noisy channel $\mathcal{N}$ can be achieved with vanishing error,
in the limit of many uses of the channel \cite{Shannon48}.
Shannon's key idea was to add redundancy to the message in order to compensate for the channel's noise.
He showed that the channel's communication capacity $\mathcal{C}(\mathcal{N})$ between two partners, called
Alice and Bob, is given by the maximal mutual information between Alice's input $a$
and Bob's output $b=\mathcal{N}(a)$, i.e.,
\begin{equation}\label{ShannonCapacity}
\mathcal{C}(\mathcal{N})=\max_a H(a{\rm:}b)\qquad\textrm{(bits/channel use)}.
\end{equation}

Quantum information theory \cite{N&C02} is a generalization of Shannon's information theory that has attracted
huge interest in the last decade, as it allows for
new potential applications, such as quantum communication and
entanglement distribution.
Quantum communication allows faithful transfer of quantum states
through a quantum noisy channel $\Lambda$. The quantum communication capacity $\mathcal{Q}(\Lambda)$ 
gives the number of qubits per channel use that can be reliably transmitted,
preserving quantum coherence.
It was shown in \cite{Schu96} that the coherent information $I(\Lambda,\rho_{A})$,
a function of Alice's input $\rho_A$ on channel $\Lambda$,
plays a crucial role in the definition of the quantum communication capacity.
The coherent information is
\begin{equation}\label{CoherentInfo}
I(\Lambda,\rho_{A})=I(\mathcal{I}\otimes\Lambda(\proj{\psi}_{RA}))=I(\rho_{RB}),
\end{equation}
where $\ket{\psi}_{RA}$ is the purification of $\rho_{A}$, $\mathcal{I}$ is the identity operator and $I(\rho_{RB})=S(B)-S(RB)$, where $S(X)$
is the von Neumann entropy of $\rho_X$.
By analogy with Shannon's theory, one would expect $\mathcal{Q}(\Lambda)$ to be
calculated by maximizing over a single use of the channel,
\begin{equation}
\mathcal{Q}^{(1)}(\Lambda)=\max_{\rho_{A}}I(\Lambda,\rho_A).
\end{equation}
Unfortunately, the quantum case is more complicated, as $\mathcal{Q}^{(1)}(\Lambda)$ is
known to be non-additive \cite{Smith07}.
The correct capacity definition \cite{Lloyd97} is,
\begin{equation}
\mathcal{Q}(\Lambda)=\lim_{n\rightarrow\infty}\frac{1}{n}\max_{\rho_{\bar{A}}}
I(\Lambda^{\otimes n},\rho_{\bar{A}}).
\label{eq:UnEDCap}
\end{equation}
Only for the restricted class of \textit{degradable channels} \cite{Devetak05},
is $\mathcal{Q}(\Lambda)$ known to be additive, i.e., $Q(\Lambda)=Q^{(1)}(\Lambda)$.
The channel $\Lambda$ is called degradable if
there exists a map $\mathcal{M}$ that transforms Bob's output $\rho_B$ into the environment state $\rho_E$, i.e., $\mathcal{M}(\rho_{B})=\rho_{E}$,
where $\rho_{E}={\rm Tr}_{RB}[\proj{\phi}_{RBE}]$ and $\ket{\phi}_{RBE}$ is the purification of $\rho_{RB}$.
Similarly if there is a map $\mathcal{G}$ such that $\mathcal{G}(\rho_{E})=\rho_{B}$ the channel
is called \textit{antidegradable} and $\mathcal{Q}(\Lambda)=0$.

\begin{figure}[!h!]
\begin{center}
\includegraphics[width=7cm]{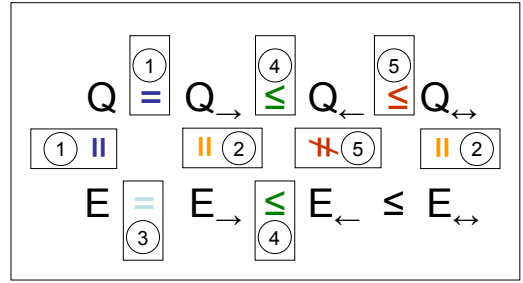}
\end{center}
\caption{(color online) Relations between the
quantum communication and entanglement distribution
capacities. We first start by two general remarks: (I) Being able to send a noiseless qubit is a stronger resource than
distributing units of entanglement (e-bits): $\mathcal{E}_x\geq \mathcal{Q}_x$ for all $x$.
(II) Increasing the complexity of the assistance cannot decrease the capacity:
$\mathcal{X}\leq \mathcal{X}_{\leftarrow}\leq \mathcal{X}_{\leftrightarrow}$.
The following remarks concern their corresponding number on the figure.
(1) The equality $\mathcal{E}=\mathcal{Q}=\mathcal{Q}_{\rightarrow}$ was shown  in \cite{Barnum00}.
(2) Any entanglement distribution protocol with free forward classical communication can be transformed
into a quantum communication protocol by appending teleportation to it.
(3) Results from combining 1 and 2.
(4) Combining 1, 3 and II.
(5) It is easy to prove that $\mathcal{E}_{\leftarrow}=\mathcal{Q}_{\leftrightarrow}$
for the erasure channel  \cite{ErasureChannel,Shor07}.
In \cite{Shor07} it was shown that the erasure channel satisfies the strict inequality $\mathcal{Q}_{\leftarrow}<\mathcal{Q}_{\leftrightarrow}$, which gives
$\mathcal{E}_{\leftarrow}\neq \mathcal{Q}_{\leftarrow}$.}
\label{Fig:CapRel}
\end{figure}

Having free access to a classical communication channel Alice and Bob can improve the
quantum communication protocol, as opposed to Shannon's theory where using feedback gives no improvement
\cite{Cover&Thomas}.
One can define three new quantum communication capacities depending on the use of the classical channel:
forward classical communication ($\mathcal{Q}_{\rightarrow}$); feedback classical communication ($\mathcal{Q}_{\leftarrow}$); two-way classical communication ($\mathcal{Q}_{\leftrightarrow}$).
In Fig.~\ref{Fig:CapRel} we review the relations between these four capacities.

Entanglement is another important resource for quantum information processing.
Therefore, the study of the entanglement distribution capacity
of quantum channels (distributed e-bits per use of the channel) is of crucial importance.
As for quantum communication, we can also define four types of assisted (unassisted) capacities for entanglement distribution:
$\{\mathcal{E},\mathcal{E}_{\rightarrow},\mathcal{E}_{\leftarrow},\mathcal{E}_{\leftrightarrow}\}$.
As shown in Fig.~\ref{Fig:CapRel}, all the entanglement distribution capacities
are equivalent to their quantum communication counterparts,
except for $\mathcal{E}_{\leftarrow}(\Lambda)$.

\begin{figure}[!h!]
\begin{center}
\includegraphics[width=8cm]{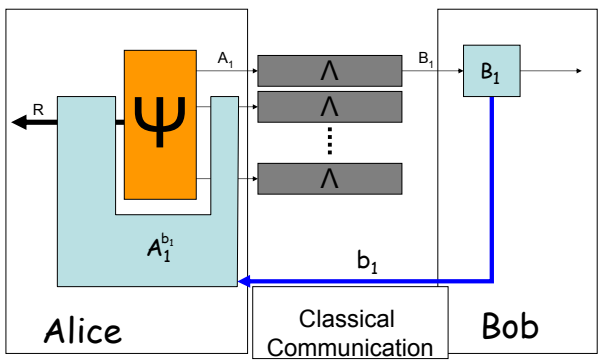}
\end{center}
\caption{(color online) The first round of the entanglement distribution protocol assisted by classical feedback
consists of three steps: i) Alice sends qubit $A_1$ through the quantum channel $\Lambda$; ii) Bob applies an incomplete quantum measurement $\mathcal{B}_1$ over his
received qubit $B_1$ and communicates the outcome $b_1$ to Alice;
iii) Alice applies a global quantum operation $\mathcal{A}_1^{b_1}$
over the joint system of $R$ and the remaining $n-1$ qubits $A_i$.
The next rounds are straightforward extensions of the first one.}
\label{Fig:EDfeedback}
\end{figure}

\textit{Entanglement distribution assisted by feedback classical communication.-}
The entanglement distribution protocol assisted by classical feedback communication,
as described in \cite{Leung08}, goes as follows.
Alice starts preparing a bipartite entangled state $\Psi_{R|A_1,A_2,...,A_n}$,
where $R$ is a group of qubits entangled with the qubits $A_i$
sent, one by one, through the channel $\Lambda$.
The first round of the protocol, see Fig.~\ref{Fig:EDfeedback}, consists of three steps: i) Alice sends qubit $A_1$ through the quantum channel $\Lambda$; ii) Bob applies an incomplete quantum measurement $\mathcal{B}_1$ over his
received qubit $B_1$ and communicates the classical outcome $b_1$ to Alice.
iii) Alice, conditioned on the classical message $b_1$, applies a global quantum operation $\mathcal{A}_1^{b_1}$
over the joint system of $R$ and the remaining $n-1$ qubits $A_2A_3...A_n$.
The next $n-1$ rounds are a slight modification of the first one: First, Bob's measurement $\mathcal{B}^{b_1...b_{i-1}}_i$ acts on all his received qubits $B_1B_2...B_i$,
conditioned on his previous measurement outcomes $b_1...b_{i-1}$.
Second, Alice's operation $\mathcal{A}^{b_1...b_{i}}_i$,
acts on all her remaining qubits $RA_{i+1}...A_n$, conditioned on all previous classical communication messages.
By properly choosing Alice's operations and Bob's incomplete measurements
both partners extract $\approx n \mathcal{E}_{\leftarrow}(\Lambda)$ units of entanglement (e-bits)
at the end of the protocol.
Unfortunately, the calculation of $\mathcal{E}_{\leftarrow}(\Lambda)$ is extremely challenging in full generality.

\begin{figure}[!h!]
\begin{center}
\includegraphics[width=8cm]{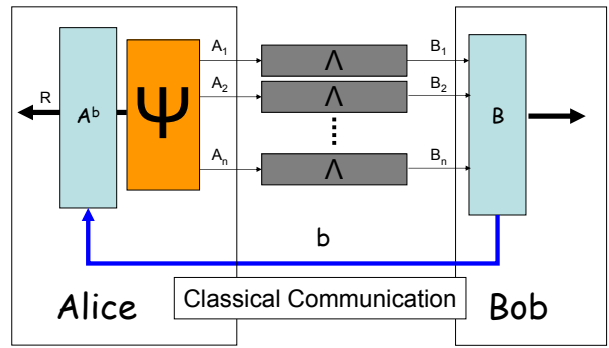}
\end{center}
\caption{(color online) A simplification of the general entanglement distribution protocol assisted
by classical feedback (Fig.\ref{Fig:EDfeedback}) limits the protocol
to a last single round of processing. After Alice has sent all her qubits ($A_1A_2...A_n$) through the quantum channel
$\Lambda$, Bob applies a collective incomplete measurement $\mathcal{B}$ among all the qubits
$B_1B_2...B_n$ and communicates
the classical outcome $b$ to Alice. Finally, conditioned on the message $b$, Alice
applies the quantum operation $\mathcal{A}^ b$ on system $R$.}
\label{Fig:RRED}
\end{figure}

\textit{Reverse entanglement distribution.-}
A big practical disadvantage of the previous protocol is that
Alice has to wait until Bob sends the message $b_i$ before applying $\mathcal{A}^{b_1...b_{i}}_i$
and subsequently sending qubit $A_{i+1}$,
which greatly decreases the transmission rate.
A way of avoiding this problem is to simplify the
protocol to a single round of classical feedback after
Alice has sent all her qubits $A_1A_2...A_n$ through the quantum channel $\Lambda$, see Fig.~\ref{Fig:RRED}.
We call this familly of simplified protocols \textit{reverse entanglement distribution protocols},
by analogy with the quantum key distribution scenario \cite{NatureCerf03}.
Before the single post-processing round Alice and Bob's shared state is
\begin{equation}
\rho_{R|B_1,B_2,...,B_n}=\mathcal{I}\otimes\Lambda^{\otimes n}(\Psi_{R|A_1,A_2,...,A_n}).
\end{equation}
By properly choosing Alice's and Bob's operations both partners extract $\approx n \mathcal{E}_{\lhd}(\Lambda)$ e-bits,
where $\mathcal{E}_{\lhd}(\Lambda)$ is the reverse entanglement distribution capacity, satisfying the inequality
$\mathcal{E}_{\lhd}(\Lambda)\leq\mathcal{E}_{\leftarrow}(\Lambda)$.

Remark that, in the particular case where Alice's inputs are independent and identically distributed, i.e.,
$\rho_{R|A_1,A_2,...,A_n}=\rho_{R|A}^{\otimes n}$, the post-processing of the reverse entanglement distribution protocol
is the dynamical equivalent of an entanglement distillation protocol over the static resource $\rho_{R|B}^{\otimes n}$ \cite{Devetak04}.

\textit{Reverse coherent information capacity.-}
In what follows we consider a subset of the reverse entanglement distribution protocols
with a strikingly simple capacity that lower bounds $\mathcal{E}_{\lhd}(\Lambda)$.
By exchanging the roles of Alice and Bob in
the family of static distillation protocol assisted by one-way classical communication defined in \cite{Devetak04},
we obtain a new family of static distillation protocols with rate
\begin{equation}
I_R(\rho_{RB})=S(R)-S(RB).
\end{equation}
By analogy with the quantum key distribution scenario \cite{NatureCerf03},
we call the quantity $I_R(\rho_{RB})$ the \emph{reverse coherent information}.
It is then straightforward to consider
a family of entanglement distribution protocols assisted by classical feedback with rate
$I_R(\Lambda,\rho_{A})=I_R(\mathcal{I}\otimes\Lambda(\proj{\psi}_{RA}))=I_R(\rho_{RB})$.
Optimizing this rate over $\rho_A$ we define the single-letter reverse coherent information capacity
$\mathcal{E}_{R}^{(1)}(\Lambda)$.

Similarly to Eq.~(\ref{eq:UnEDCap}) we can define a regularized entanglement capacity $\mathcal{E}_{R}(\Lambda)$
that lowerbounds $\mathcal{E}_{\lhd}(\Lambda)$.
Interestingly, this quantity can be shown to be additive for all channels, i.e., $\mathcal{E}_{R}=\mathcal{E}_{R}^{(1)}$.
To do so we only need to prove the relation
\begin{equation}
I_R(\Lambda\otimes\Lambda,\rho_{A_1A_2})\leq I_R(\Lambda,\rho_{A_1})+I_R(\Lambda,\rho_{A_2}).
\label{eq:addRRED}
\end{equation}
Using the alternative definition of the reverse coherent information
$I_R(\rho_{RB})=S(BE)-S(E)$, where $\ket{\phi}_{RBE}$
is the purification of $\rho_{RB}$ and $\rho_{BE}$, Eq.~(\ref{eq:addRRED})
can be restated as a relation between two von Neumann mutual information quantities:
$S(B_1E_1{\rm :}B_2E_2)\geq S(E_1{\rm :}E_2)$. This relation holds because
discarding quantum systems can only decrease the mutual information, which
results from the strong-subadditivity of the entropy.

The previous proof is strikingly similar to the additivity
of the unassisted capacity of degradable channels,
except that it holds for all channels.
Since $I_R(\Lambda,\rho_{A})$ is additive, it would be extremely interesting if it could
be used to give a definition of $\mathcal{E}_{\leftarrow}(\Lambda)$ or
$\mathcal{E}_{\lhd}(\Lambda)$ similar to Eq.~(\ref{eq:UnEDCap}).
Unfortunately, this cannot be done as $I_R(\Lambda,\rho_{A})$ does not satisfy the
data processing inequality.

Despite reverse coherent information capacity restricts
the protocols to a very specific subset,
its study remains very interesting,
as for some channels it achieves a remarkable improvements over the unassisted capacity $\mathcal{E}(\Lambda)$.
To get some intuition on when we may obtain an improvement, we look at the difference
between the coherent information and its reverse counterpart
($I_{R}(\rho_{RB})-I(\rho_{RB})=S(R)-S(B)$).
We see that for channels satisfying $S(R)>S(B)$ over all inputs,
such as the bosonic lossy channel, reverse reconciliation performs better than $\mathcal{E}^{(1)}$.
On the other hand, for those channels satisfying $S(B)\geq S(R)$ for all inputs, such
as optical amplifiers or the erasure channel, we obtain $\mathcal{E}\geq\mathcal{E}^{(1)}\geq\mathcal{E}_{R}$.
In the case of the erasure channels is it easy to see that
$\mathcal{E}_{\leftrightarrow}=\mathcal{E}_{\lhd}>\mathcal{E}>\mathcal{E}_{R}$, which
gives an example of strict separation between $\mathcal{E}_{\lhd}$ and $\mathcal{E}_{R}$.

\begin{figure}[!h!]
\begin{center}
\includegraphics[width=8cm]{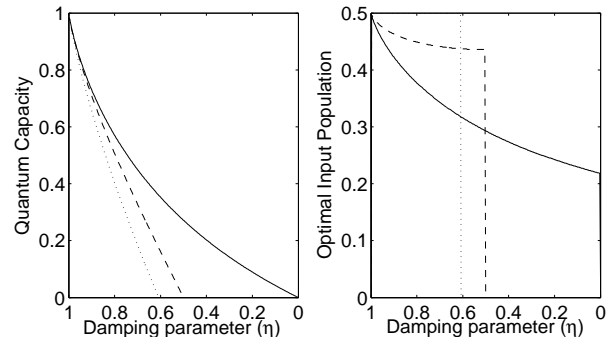}
\end{center}
\caption{(a) Comparison of $\mathcal{E}(\mathcal{D}_{\eta})$ (dashed line) and
$\mathcal{E}_{R}(\mathcal{D}_{\eta})$ (solid line)
as functions of the damping parameter $\eta$ for the amplitude damping channel, together with
the capacity $\mathcal{E}^{(1)}(\mathcal{D}_{\eta})=\mathcal{E}_{R}(\mathcal{D}_{\eta})$ of the generalized amplitude damping channel with a maximally mixed environment ($\alpha=1/2$)
(dotted line). (b) Optimal input population $p$ achieving the previous capacities.}
\label{Fig:ADCh}
\end{figure}

\textit{Amplitude damping channel.-}
The amplitude damping channel describes the process of energy dissipation through spontaneous emission
in a two-level system. The effect of the channel on the input state $\rho$ is
$\mathcal{D}_{\eta}(\rho)=E_0\rho E^{\dagger}_0+E_1\rho E^{\dagger}_1$, where
\begin{equation}
E_0=\left[
\begin{array}{cc}
1 & 0 \\
0 & \sqrt{\eta}
\end{array}
\right],
E_1=\left[
\begin{array}{cc}
0 & \sqrt{1-\eta}\\
0 & 0
\end{array}
\right],
\end{equation}
and $1-\eta$ is the probability of spontaneous emission.
Generalizing the results of \cite{Giovannetti05}, we can
restrict the input state to the class $\rho_{A}={\rm diag}(1-p,p)$ without
loss of generality (see appendix). For a given input population $p$,
the output state is $\rho_{B}={\rm diag}(1-\eta p,\eta p)$ and the (reverse)
coherent information becomes
\begin{eqnarray}
I(\mathcal{E},p)&=&H(\eta p)-H((1-\eta)p), \nonumber \\
I_{R}(\mathcal{E},p)&=&H(p)-H((1-\eta)p),
\end{eqnarray}
where $H(x)$ is the binary entropy.
Optimizing over the input population we obtain $\mathcal{E}^{(1)}(\mathcal{D}_{\eta})$
and $\mathcal{E}_{R}(\mathcal{D}_{\eta})$
as functions of the damping parameter $\eta$, see Fig.~\ref{Fig:ADCh}.
Using the concatenation property of the amplitude damping channel
($\mathcal{D}_{\eta}\circ\mathcal{D}_{\eta'}=\mathcal{D}_{\eta\eta'}$)
it is easy to prove that the
amplitude damping channel is degradable ($\mathcal{E}(\mathcal{D}_{\eta})=\mathcal{E}^{(1)}(\mathcal{D}_{\eta})$)
for $\eta\geq1/2$ and
antidegradable ($\mathcal{E}(\mathcal{D}_{\eta})=0$) for $\eta\leq1/2$.
We conclude that $\mathcal{E}_{R}(\mathcal{D}_{\eta})$ outperforms $\mathcal{E}(\mathcal{D}_{\eta})$ for all $\eta$. Even
more interestingly, $\mathcal{E}_{R}(\mathcal{D}_{\eta})$ remains positive in the range $\eta\leq1/2$ where $\mathcal{E}(\mathcal{D}_{\eta})=0$, see Fig.~\ref{Fig:ADCh} (a).

\begin{figure}[!h!]
\begin{center}
\includegraphics[width=7cm]{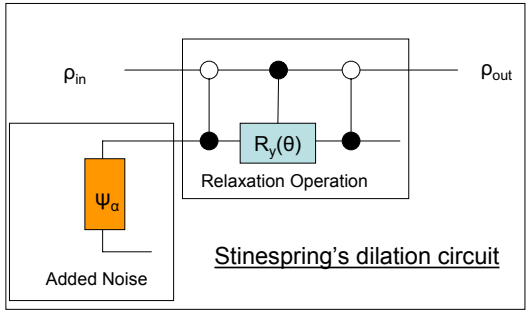}
\end{center}
\caption{(color online) Quantum circuit corresponding to the Stinespring's dilation of
the generalized amplitude damping channel $\mathcal{D}_{(\eta,\alpha)}$.
Alice's input state $\rho_{in}$ and half of an entangled state $\ket{\Psi_{\alpha}}$ interact through
the relaxation operation $U_{{\rm RO}}$ composed of two CNOT gates and a controlled rotation
around the $y$-axis of the Bloch sphere ($\cos^2(\gamma/2)=\eta$).}
\label{Fig:CircuitGADCh}
\end{figure}

\textit{Generalized amplitude damping channel.-}
Spontaneous emission to
an environment at thermal equilibrium leads to the generalized amplitude
damping channel $\mathcal{D}_{(\eta,\alpha)}$, which can be modeled by the Stinespring's dilation circuit
of Fig.~\ref{Fig:CircuitGADCh}.
The relaxation operation applies the unitary transformation,
\begin{equation}
U_{{\rm RO}}=\left[
\begin{array}{cccc}
1 & 0 & 0 & 0 \\
0 & \sqrt{\eta} & \sqrt{1-\eta} & 0 \\
0 & -\sqrt{1-\eta} & \sqrt{\eta} & 0 \\
0 & 0 & 0 & 1
\end{array}
\right],
\end{equation}
jointly to the input state and the environment.
The thermal environment is modeled by inserting half of an
entangled state $\ket{\Psi_{\alpha}}=\sqrt{1-\alpha}\ket{00}+\sqrt{\alpha}\ket{11}$ into
the second input of $U_{{\rm RO}}$.
The channel can be seen as the random mixing
$\mathcal{D}_{(\eta,\alpha)}=\alpha\mathcal{D}_{(\eta,0)}+(1-\alpha)\mathcal{D}_{(\eta,1)}$
of two limiting cases: (1) the amplitude damping channel when  ($\mathcal{D}_{(\eta,0)}$); and (2)
a \emph{populating channel} ($\mathcal{D}_{(\eta,1)}$).
We restrict the analysis to $0\leq\alpha\leq1/2$ as for
any channel $\mathcal{D}_{(\eta,\alpha=1/2+x)}$ with optimal input population $p^*$
there is a symmetric channel $\mathcal{D}_{(\eta,\alpha=1/2-x)}$  with optimal population $1-p^*$
reaching the same capacity.
As before, Alice's input can be restricted to $\rho_{A}={\rm diag}(1-p,p)$ (see appendix) giving
\begin{eqnarray}
S(B)&=&H(\eta p+(1-\eta)\alpha), \\
S(AB)&=&H_4(\lambda_1,\lambda_2,\lambda_3,\lambda_4),
\end{eqnarray}
where $H_4$ is the Shannon entropy of a $4$-dimensional distribution
and $\lambda_j$ are the four eigenvalues of $\rho_{AB}$,
\begin{eqnarray}
\lambda_1&=&\alpha(1-\eta)(1-p), \lambda_2=(1-\alpha)(1-\eta)p, \\
\lambda_{3,4}&=&\left[1-\lambda_1-\lambda_2
\pm\sqrt{1-2(\lambda_1+\lambda_2)+(\lambda_2-\lambda_1)^2}\right]/2 \nonumber.
\end{eqnarray}
\begin{figure}[!h!]
\begin{center}
\includegraphics[width=7.5cm]{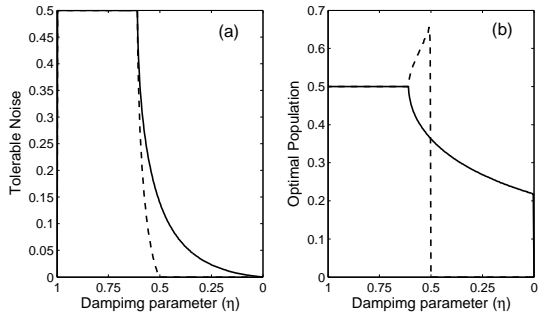}
\end{center}
\caption{(a) Tolerable thermal noise of the generalized amplitude damping channel $\mathcal{D}_{(\eta,\alpha)}$
(minimum $\alpha$ such that the capacity is zero)
as a function of the damping parameter $\eta$ for: $\mathcal{E}^{(1)}(\mathcal{D}_{(\eta,\alpha)})$ (dashed line), and
$\mathcal{E}_{R}(\mathcal{D}_{(\eta,\alpha)})$ (solid line). (b) Input population $p$ achieving the curves of (a).}
\label{Fig:PlotGADCh}
\end{figure}
Optimizing over the input population $p$ we obtain the capacities
$\mathcal{E}^{(1)}(\mathcal{D}_{(\eta,\alpha)})$ and $\mathcal{E}_{R}(\mathcal{D}_{(\eta,\alpha)})$.
It is easy to show that
$\mathcal{E}_{R}(\mathcal{D}_{(\eta,\alpha)})>\mathcal{E}^{(1)}(\mathcal{D}_{(\eta,\alpha)})$ for
any noise $\alpha$ except for $\alpha=1/2$, where both are equal, as shown in
Fig.~\ref{Fig:ADCh}. Unfortunately, we cannot conclude
$\mathcal{E}_{R}(\mathcal{D}_{\eta,\alpha})\geq \mathcal{E}(\mathcal{D}_{(\eta,\alpha)})$ for $\alpha>0$,
as the channels are no longer degradable.
Nevertheless,
it is easy to prove that generalized amplitude damping channels ($\mathcal{D}_{(\eta,\alpha)}$) with $\eta\leq1/2$ are
antidegradable ($\rho_{B}=\mathcal{D}_{(\eta/(1-\eta),\alpha)}(\rho_E)$), which shows that for such channels
$\mathcal{E}_{R}(\mathcal{D}_{(\eta,\alpha)})\geq \mathcal{E}(\mathcal{D}_{(\eta,\alpha)})=0$
(see Fig.~\ref{Fig:PlotGADCh}).

\textit{Conclusion.-}
We reviewed the relation between quantum communication and entanglement distribution capacities,
paying special attention to entanglement distribution assisted by classical feedback.
By restricting ourselves to realistic protocols with a single final round of post-processing, we
defined the reverse entanglement distribution protocols. A subset of such protocols
give an operational interpretation of the reverse coherent information, a symmetric counterpart of the coherent information.
This allow us to define a new entanglement distribution capacity
which is additive and outperforms the unassisted capacity
for some important channels, such as the damping channel and its generalization.



We acknowledge financial support from the W. M. Keck Foundation Center for Extreme Quantum Information Theory.
S.P. acknowledges financial support from the EU (Marie Curie fellowship).

\section*{Appendix: Optimality of the input state}

In this appendix we show that the input state $\rho_{A}={\rm diag}(1-p,p)$ 
maximizes the (reverse) coherent information of the amplitude damping channel and its generalization.
The coherent information for degradable channels (amplitude damping)
being a concave function implies that diagonal input states outperform non-diagonal states \cite{Perez07}. 
The same argument hold for the reverse
coherent information, this time over all channels. 

The optimization of the coherent information for non-degradable channels, such as
the generalized amplitude damping channels, needs a more detailed proof.
For shake of completeness we present this specific proof for the
(reverse) coherent information for 
all channels studied in this manuscript.

\subsection{Amplitude Damping Channel}

\begin{figure}[!t!]
\begin{center}
\includegraphics[width=6cm]{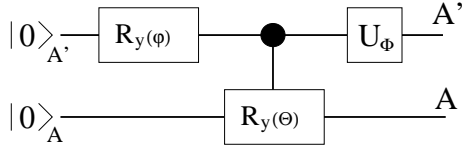}
\end{center}
\caption{Quantum circuit generating the bipartite state $\ket{\psi}_{AA'}$. The first rotation
$R_y(\varphi)$ generates the quantum state $\sqrt{1-p}\ket{0}_{A'}+\sqrt{p}\ket{1}_{A'}$ ($\sin(\varphi/2)=p$).
The bipartite state is then entangled by the controlled rotation, generating $\ket{\psi^{(\phi=0)}}_{AA'}$. 
The phase $\phi$ is finally fixed by a last local unitary operation $U_{\phi}$ on $A'$.}
\label{Fig:CircuitApp}
\end{figure}

The most general input state to the amplitude damping (AD) channel reads
\begin{equation}
\rho_{A'}=\left[
\begin{array}{cc}
1-p & \sqrt{(1-p)p}e^{-i\phi}\cos\theta \\
\sqrt{(1-p)p}e^{i\phi}\cos\theta & p
\end{array}
\right].
\end{equation}
One valid purification of $\rho_{A'}$ (all are equivalent up to a unitary on $A$) reads
\begin{equation}
\ket{\psi}_{AA'}=\sqrt{1-p}\ket{0}_A\ket{0}_{A'}+\sqrt{p}e^{i\phi}\ket{1}_{A'}[\cos\theta\ket{0}+\sin\theta\ket{1}]_{A}.
\end{equation}
In Fig.~\ref{Fig:CircuitApp} we observe that the state $\ket{\psi^{\phi}}_{AA'}$ 
is generated from $\ket{\psi^{(\phi=0)}}_{AA'}$  by the local unitary 
$
U_{\phi}=\mathbb{I}_A\otimes\left[
\begin{array}{cc}
1 & 0 \\
0 & e^{-i\phi}
\end{array}
\right]_{A'}
$
applied just before sending the state $A'$ through the channel. 

After passage through the channel Alice and Bob entangled state reads
\begin{widetext}

\begin{equation}
\rho_{AB}(\phi)=\left[
\begin{array}{cccc}
1-p+p(1-\eta)\cos^2\theta & p(1-\eta)\cos\theta\sin\theta & \sqrt{\eta(1-p)p}e^{-i\phi}\cos\theta & \sqrt{\eta(1-p)p}e^{-i\phi}\sin\theta \\
p(1-\eta)\cos\theta\sin\theta & p(1-\eta)\sin^2\theta & 0 & 0 \\
\sqrt{\eta(1-p)p}e^{i\phi}\cos\theta & 0 & p\eta\cos^2\theta & p\eta\cos\theta\sin\theta \\
\sqrt{\eta(1-p)p}e^{i\phi}\sin\theta & 0 & p\eta\cos\theta\sin\theta & p\eta\sin^2\theta
\end{array}
\right],
\end{equation}
\end{widetext}
where $\eta$ is the damping parameter.
It is easy to check that by applying $U_{\phi}^{\dagger}$ to $\rho^{\phi}_{AB}$
we obtain $\rho^{(\phi=0)}_{AB}$. 
Because the von Neumann entropy is invariant under a unitary transformation $U_{\phi}^{\dagger}$, 
we can restrict our study to the case $\phi=0$ without loss of generality.

\subsubsection{Eigenvalues}

The eigenvalues of $\rho_{A}$ read
\begin{equation}
\lambda_{A,1(2)}=\lambda_{\pm}(1).
\end{equation}
where
\begin{equation}
\lambda_{\pm}(x)=[1\pm\sqrt{(1-2xp)^2+4x(1-p)p\cos^2\theta}]/2.
\end{equation}

The eigenvalues of $\rho_{AB}$ then read
\begin{eqnarray}
\lambda_{AB,1(2)}&=&\lambda_{\pm}(1-\eta), \nonumber \\
\lambda_{AB,3(4)}&=&0.
\end{eqnarray}
Bob state read
\begin{equation}
\rho_{B}=\left[
\begin{array}{cc}
1-\eta p & \sqrt{\eta(1-p)p}\cos\theta \\
\sqrt{\eta(1-p)p}\cos\theta & \eta p
\end{array}
\right],
\end{equation}
which gives the eigenvalues
\begin{equation}
\lambda_{B,1(2)}=\lambda_{\pm}(\eta).
\end{equation}
Now we are ready to calculate the (reverse) coherent information for a general input state.

\subsubsection{Coherent Information}
The coherent information reads $I=S(B)-S(AB)$. In order to proof that $\cos\theta=0$
maximizes the coherent information, we calculate the derivative of $I$,
\begin{eqnarray}
\frac{\partial I}{\partial\theta}&=&-p(1-p)\sin(2\theta)
\left[F(\eta)-F(1-\eta)\right]
\end{eqnarray}
where
\begin{equation}
F(x)=\frac{x}{\sqrt{a}}\log\left[\frac{1-\sqrt{a}}{1+\sqrt{a}}\right],
\end{equation}
and $a=(1-2x p)^2+4x(1-p)p\cos^2\theta$.
In order to find the values of $\theta$ maximizing $I$, we first search for the extrema
($\partial I/\partial\theta=0$). The term $\sin(2\theta)$ ($\cos^2\theta$) having period of $\pi$ ($\pi/2$), 
we can restrict the study to the domain $\theta\in\{0,\pi\}$ without loss of generality.
The are two cases of pathological extrema; Firstly, $p=0$ and $p=1$ which correspond to
separable input states ($\ket{00}$ and $\ket{11}$ respectively) which give $I=0$; 
Secondly, $F(\eta)=F(1-\eta)$ giving $\eta=1/2$, i.e., the range limitation of $I$ 
(the lowest $\eta$ such that $I(\eta=1/2)=0$). For $0<p<1$ and $\eta>1/2$ we have an extremum when $\sin(2\theta)=0$,
($\theta=k\pi/2$). 
For $\theta=\{0,\pi\}$ the input state is separable ($\ket{0}_A\otimes(\ket{0}\pm\ket{1})_{A'}/\sqrt{2}$) 
and therefore has $I=0$ (as $S(A,B)=S(A)+S(B)$ and $S(A)=0$), which is a minimum for $\eta>1/2$.
Because a single extremum between two minimums can only be a maximum, we conclude that
$\theta=\pi/2$ optimizes $I$. We have then proven
that the optimal input is ${\rm diag}(p,1-p)$, as
shown in \cite{Giovannetti05}. In the range $\eta<1/2$ the roles of $\pi/2$ and $\theta=\{0,\pi\}$ are exchanged,
giving $I=0$ as maximum.

\subsubsection{Reverse Coherent Information}
The reverse coherent information reads $I_{R}=S(A)-S(AB)$. The proof is very similar to the previous result,
where the partial derivative among $\theta$ now reads,
\begin{equation}
\frac{\partial I}{\partial\theta}=-p(1-p)\sin(2\theta)\left[F(1)-F(1-\eta)\right].
\end{equation}
For $\theta=\{0, \pi\}$
the initial input being separable, we obtain
$I_R=-S(B)$, which is negative. This two extrema being minima, 
$\theta=\pi/2$ remains the maximum. 
The pathological extremum
$\eta=1/2$ is now replaced by $\eta=0$, which coincides with the range limitation of the reverse coherent information.

\subsection{Generalized Amplitude Damping Channel}
After passage through the generalized amplitude damping (GAD) channel, Alice and Bob entangled state reads
\begin{widetext}
\begin{equation}
\rho_{AB}=\left[
\begin{array}{cc}
Z & C \\
C^T & W
\end{array}
\right],
\end{equation}
where
\begin{equation}
Z=\left[
\begin{array}{cc}
(1-p)\alpha\eta+(1-\alpha)(1-p+p(1-\eta)\cos^2\theta) & p(1-\alpha)(1-\eta)\cos\theta\sin\theta  \\
p(1-\alpha)(1-\eta)\cos\theta\sin\theta  & p(1-\alpha)(1-\eta)\sin^2\theta
\end{array}
\right],
\end{equation}
\begin{equation}
W=\left[
\begin{array}{cc}
\alpha(1-p)(1-\eta)+p(\alpha+(1-\alpha)\eta)\cos^2\theta & p(\alpha+(1-\alpha)\eta)\cos\theta\sin\theta \\
p(\alpha+(1-\alpha)\eta)\cos\theta\sin\theta & p(\alpha+(1-\alpha)\eta)\sin^2\theta,
\end{array}
\right],
\end{equation}
\begin{equation}
C=\left[
\begin{array}{cc}
\sqrt{\eta(1-p)p}\cos\theta & \sqrt{\eta(1-p)p}\sin\theta \\
0 & 0
\end{array}
\right],
\end{equation}
$\eta$ is the damping parameter and $\alpha$ is related to the thermal noise of the environment.
After a long calculation one can show that the eigenvalues of $\rho_{AB}$ read
\begin{eqnarray}
\lambda_{AB,1(2)}&=&\frac{1}{4}\left[1+\sqrt{a+bp(1-p)\cos^2\theta}\pm\sqrt{c+dp(1-p)\cos^2\theta+\sqrt{a+bp(1-p)\cos^2\theta}}\right], \nonumber \\
\lambda_{AB,3(4)}&=&\frac{1}{4}\left[1-\sqrt{a+bp(1-p)\cos^2\theta}\pm\sqrt{c+dp(1-p)\cos^2\theta-\sqrt{a+bp(1-p)\cos^2\theta}}\right].
\end{eqnarray}
where
\begin{eqnarray}
a&=&\left(1-2(1-\eta)(\alpha+p-2\alpha p)\right)^2 \\
b&=&4(1-\eta)(1-4(1-\alpha)\alpha(1-\eta)) \\
c&=&1-2(1-\eta)(\alpha+p-2\alpha p)+2(p-\alpha)^2(1-\eta)^2 \\
d&=&2(1-\eta).
\end{eqnarray}
\end{widetext}
In the GAD channel Alice state $\rho_A$ and its eigenvalues remain
the same as in the AD channel. Bob's state reads
\begin{equation}
\rho_{B}=\left[
\begin{array}{cc}
1-(\eta p+(1-\eta)\alpha) & \sqrt{\eta(1-p)p}\cos\theta \\
\sqrt{\eta(1-p)p}\cos\theta & \eta p+(1-\eta)\alpha
\end{array}
\right],
\end{equation}
which gives the eigenvalues
\begin{equation}
\lambda_{B,1(2)}=\frac{1}{2}\left[1\pm\sqrt{e+fp(1-p)\cos^2\theta}\right],
\end{equation}
where $e=(1-2(p\eta+\alpha(1-\eta)))^2$ and $f=4\eta$.
\subsubsection{Extremum Search}
The first derivative of the coherent information reads,
\begin{equation}
\frac{\partial I}{\partial\theta}=\frac{p(1-p)}{8}\sin(2\theta)
\left[Z+\sum_{i,j=0}^1Y(i,j)\left[1+J(i,j)\right]\right],
\label{eq:GAD-deriv}
\end{equation}

\begin{widetext}
where,
\begin{equation}
Z=-\frac{f}{\sqrt{e+fp(1-p)\cos^2\theta}}
\log\left[\frac{1-\sqrt{e+fp(1-p)\cos^2\theta}}{1+\sqrt{e+fp(1-p)\cos^2\theta}}\right],
\end{equation}
\begin{equation}
Y(i,j)=(-)^i\left[\frac{b}{\sqrt{a+b\cos^2\theta}}+(-)^j\frac{\frac{b}{\sqrt{a+b\cos^2\theta}}-(-)^i2d}
{\sqrt{2}\sqrt{c+d\cos^2\theta-(-)^i\sqrt{a+b\cos^2\theta}}}\right].
\end{equation}
and
\begin{equation}
J(i,j)=\log\left[\frac{1}{4}\left(1-(-)^i\sqrt{a+bp(1-p)\cos^2\theta}+(-)^j\sqrt{2}
\sqrt{c+dp(1-p)\cos^2\theta-(-)^i\sqrt{a+bp(1-p)\cos^2\theta}}\right)\right]. 
\end{equation}
\end{widetext}
In order to find the maxima of $I$ as a function of $\theta$, we search first for the extrema of $I$
on the domain $\theta\in\{0,\pi\}$. As for the AD channel,  $p=0$ and $p=1$ are extrema. 
The term $\sin(2\theta)$ gives us again two sets of extrema; Firstly, $\theta=\{0,\pi\}$ 
corresponding to unentangled inputs giving $I=0$;
Secondly, the solution $\theta=\pi/2$, corresponding to the conjectured optimal input, which is the candidate 
for being the maximum. Unfortunately the complicated form of the solution of eq.~(\ref{eq:GAD-deriv})
does not preclude the existence of new extrema. 
\begin{figure}[!h!]
\begin{center}
\includegraphics[width=8cm]{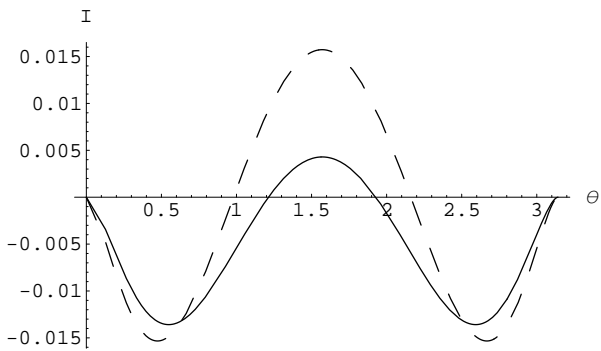}
\end{center}
\caption{Coherent information $I$ as a function of $\theta$ 
for the GAD channel with parameters $\eta=0.62$, $\alpha=0.5$, and input population
$p=0.25$ (solid line) and $p=0.5$ (dashed line).}
\label{Fig:AppFig}
\end{figure}
After carrying a detailed numerical 
check over a large spectra of values of the parameter $p,\alpha$ and $\eta$ we have seen that there exist
always two extra solutions to the equation $\partial I/\partial\theta$=0, as shown in Fig.~\ref{Fig:AppFig},
except for $\alpha=0$, i.e., the AD channel.
The numerical check
shows that both solutions $\{\theta^*,\pi-\theta^*\}$ are always minima. 
Therefore we conclude that $\theta=\pi/2$ is the maximum, as expected.

\subsubsection{Reverse Reconciliation}

\begin{figure}[!h!]
\begin{center}
\includegraphics[width=8cm]{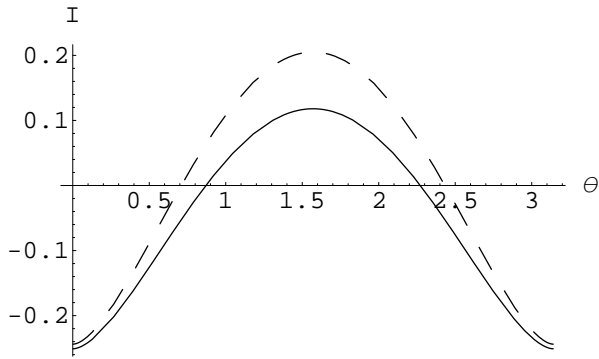}
\end{center}
\caption{Reverse coherent information $I_R$ as a function of $\theta$ 
for the GAD channel with parameters $\eta=0.75$, $\alpha=0.4$, and input population
$p=0.25$ (solid line) and $p=0.5$ (dashed line).}
\label{Fig:AppFig2}
\end{figure}

In the case of reverse coherent information the proof is very similar,
we just need to change $e$ and $f$ by $e'=(1-2p)^2$ and $f'=4$. 
The maxima and minima remains the same than those of the coherent information,
except for the two minima $\{\theta^*,\pi-\theta^*\}$ which no longer exist, as
shown in Fig.~\ref{Fig:AppFig2}.
We also observe that for $\theta=\{0,\pi\}$ the reverse coherent information
is negative ($I_R=-S(B)$), as for the AD channel. 

\end{document}